\documentclass[twocolumn,showpacs,prl,superscriptaddress]{revtex4-1}
 \pdfoutput=1

\usepackage{amsmath}
\usepackage{epsf}
\usepackage{graphicx}
\usepackage{dcolumn}
\usepackage{bm}

\newcommand{\bpsi}{\mbox{\boldmath $\psi$}}

\newcommand{\req}[1]{Eq.~(\ref{#1})}
\newcommand{\reqs}[1]{Eqs.~(\ref{#1})}
\newcommand{\rref}[1]{(\ref{#1})}

\newcommand{\beq}{\begin{equation}}
\newcommand{\eeq}{\end{equation}}
\newcommand{\be}{\begin{equation}}
\newcommand{\ee}{\end{equation}}
\newcommand{\beqa}{\begin{eqnarray}}
\newcommand{\eeqa}{\end{eqnarray}}
\newcommand{\bea}{\begin{eqnarray}}
\newcommand{\eea}{\end{eqnarray}}
\usepackage{amssymb}
\usepackage{array}
\usepackage{amsmath}
\usepackage{graphicx}\usepackage{graphics}
\usepackage{dcolumn}
\usepackage{bm}\usepackage{varioref}
\newcommand{\hH}{\hat{H}}\newcommand{\hS}{\hat{
S}}
\newcommand{\hO}{\hat{O}}

\newcommand{\htau}{\hat{\tau}}

\begin{document}

\title{Localization at the Edge of 2D Topological Insulator by Kondo Impurities with Random Anisotropies}

\author{B.L. Altshuler}
\author{ I.L. Aleiner}
\affiliation{Physics Department, Columbia University, New York, 10027, USA}
\author{V.I. Yudson}
\affiliation{Institute for Spectroscopy, Russian Academy of Sciences, Troitsk, Moscow
142190, Russia}


\begin{abstract}
We consider chiral electrons moving along the 1D helical 
edge of a 2D topological insulator and interacting with 
a disordered chain of Kondo impurities. Assuming the 
electron-spin couplings of random anisotropies, we map 
this system to the  problem of the pinning of 
the charge density wave by the disordered potential.
This mapping proves that arbitrary weak anisotropic disorder in 
coupling of chiral electrons with spin impurities leads to the
Anderson localization of the edge states.

\end{abstract}

\maketitle

{\em Introduction.} -- 
Recent interest to the topological insulators (TI) is inspired by remarkable properties of their boundaries \cite{KaneMele,Molenkamp,Review,Yacoby}. While the charged excitations in the bulk of TI are gapped as they are in conventional band insulators, the boundary can host gapless excitations. In the presence of a potential disorder there appear bulk electronic states in the gap. However such states are localized and thus cannot support any DC current. At the same time the boundary states of TI remain extended making the system conductive. The prediction is most dramatic for the one-dimensional (1D) edge of a two-dimensional (2D) TI where right and left moving electrons carry opposite spins: the conductance remains perfect ($e^2/h$) because the potential disorder cannot flip spins of the edge electrons and thus cannot back-scatter them. As a result the usual 1D Anderson localization does not occur. On the other hand, surface imperfections in real crystals are by no means limited by the potential disorder. Since the existing experiments \cite{Molenkamp,Yacoby} do not show perfect conductance even for short TI edges, it is important to understand the possible sources of the backscattering.

\begin{figure}[h]
\includegraphics[width=0.9\columnwidth]{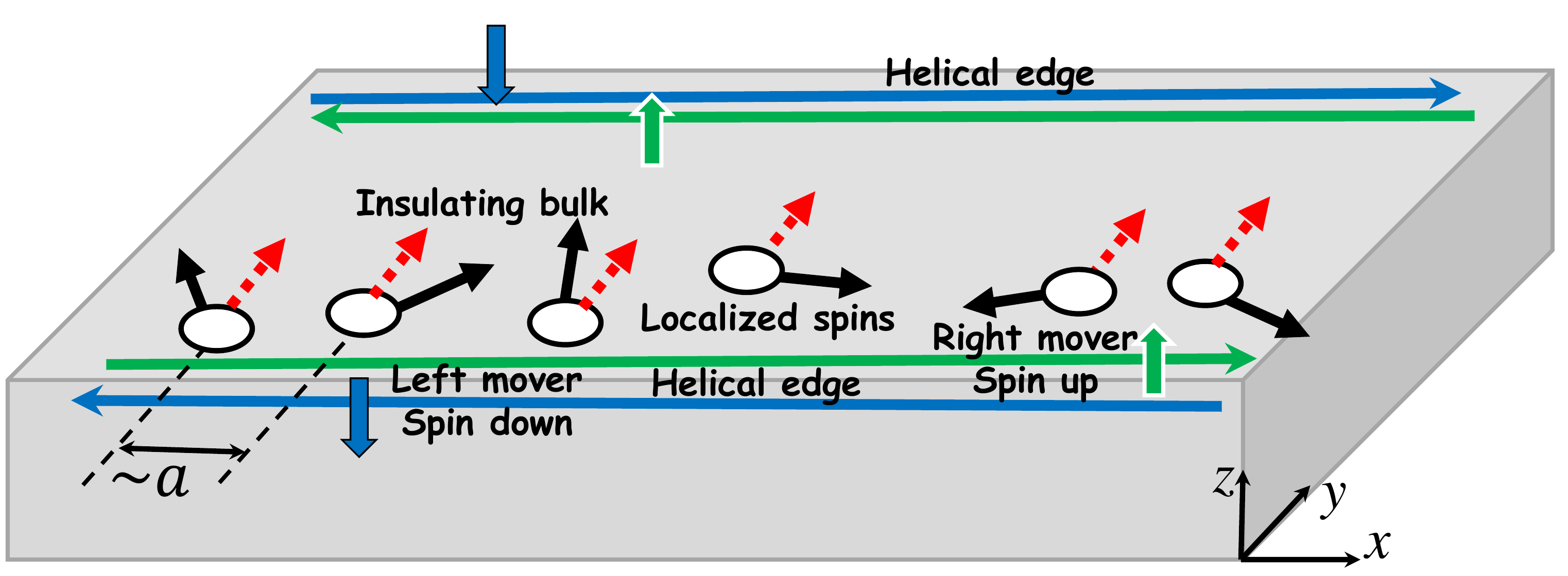}
\caption{Chiral electrons at the helical edge of a 2D TI 
 interacting with localized spins. Solid arrows 
show the original impurity spins while the dashed arrows 
illustrate the tendency to ferromagnetic in-plane ordering of
the impurity spins modified by the local rotations \rref{rotation1}.  
In combination with this ordering, the random magnetic anisotropy 
leads to the backscattering of the edge electrons and thus to 
the destruction of the chirality due to the Anderson localization.
}
\label{fig1}
\end{figure}

The scattering of the edge electrons by localized spins as shown in Fig.~\ref{fig1} seems to be the most dangerous. Such magnetic impurities appear, e.g. due to the Hubbard-like repulsion, which prevents double occupation of the localized electronic states. Indeed, the elementary process of spin exchange between the edge electron and impurities ($e-s$ coupling) is accompanied by the electron backscattering. In other words, the exchange field of the static magnetic impurities violates the time reversal symmetry and the edge states loose the chirality. However this symmetry can be restored by dynamics of the localized spins as happens e.g. in Kondo effect \cite{Kondo}. Moreover, it was shown \cite{Matveev} that the conductance should remain perfect even at the level of Boltzmann equation as long as a component of the total spin of electrons and impurities $S^{tot}_z$ is conserved, i.e., the system is invariant under rotation of all spins around $z$-axis.
On the other hand, there is no reason for the $e-s$ coupling to be isotropic in disordered systems with strong spin-orbit interaction. Some processes that violate $S^{tot}_z$ - conservation were recently discussed \cite{Glazman,Majeiko-12,Other} and their effect was argued to vanish at $T \rightarrow 0$ \cite{Glazman} or at weak enough interaction  \cite{Majeiko-12,Other}.

Here we study the effect of random anisotropy of the electron coupling with spin impurities, which breaks the conservation of $S^{tot}_z$. We show that in fact this coupling leads to backscattering which survives even at $T \rightarrow 0$ limit. Thus an arbitrary weak anisotropy of the electron-spin coupling leads to Anderson localization of edge states in long enough samples.
In spite of the fact that in real systems the spin-impurities can be distributed all over the 2D bulk, we describe them by an effective 1D model. This model is motivated by the exponential decay of the exchange interaction with the distance. Taking into account the remote spin impurities can only change initial parameters of the effective model without affecting  qualitative conclusions.

\begin{subequations}
\label{problem}
{\em Model and results.--}
The helical edge is described by the Hamiltonian
\be
\hH=-iv\int dx \bpsi^\dagger\htau_z\partial_x\bpsi+\sum_{k}\!
J_{\alpha\beta}^{k}\hS_{\alpha}^k\left[\bpsi^\dagger\htau_\beta\bpsi\right]_{x_k}
,\label{H-electron-free}
\ee
 where $v$ is the Fermi velocity and $\bpsi^\dagger,\bpsi$ are two component fermionic fields
$\bpsi^T=\left(\psi_{R,\uparrow},\psi_{L,\downarrow}\right)$. Chirality of the edge is reflected in the fact that the
 direction of the electron motion (L,R) is glued to their spin direction ($\uparrow,\downarrow $).
 Matrices $\htau_{x,y,z}$ are the Pauli matrices acting in two-dimensional spin space,
and summation over repeated indices $\alpha,\beta=x,y,z$ is always implied.
Spin $1/2$ operators $\hS_{x,y,z}^k$
describe localized spins interacting with electrons at random points $x_k$, $x_{k+1}>x_k$,
located at the typical distance $a=\langle\langle x_{k+1}-x_{k}\rangle\rangle$ from each other.
Hereinafter $\langle\langle\dots\rangle\rangle$ stands for the disorder average.

The disorder in the problem is encoded into coupling
$J_{\alpha\beta}^k$ constants. We parameterize them as
\be
\begin{split}
\hat{J}^{k}=J\hO^{k}
\begin{pmatrix}
1+\epsilon^{k} \cos\phi^k & \epsilon^{k} \sin\phi^k & j_x^k\\
\epsilon^{k} \sin\phi^k & 1-\epsilon^{k} \cos\phi^k &j_y^k\\
0 & 0 & j_z^k
\end{pmatrix}
\end{split}
\label{couplings}
\ee
where $\hO^{k}$ are orthogonal $3\times 3$ matrices \cite{Parametrization}.
Constant $J\ll v$ characterizes the typical
strength of the interaction of chiral electrons with localized spins.
The fluctuations of this overall scale do not lead to any new physics and we will
neglect them from the very beginning.
Non-vanishing $\epsilon^k$ are of the most importance  and we
adopt 
\be
\langle\langle\epsilon^{k}\epsilon^{l}e^{i(\phi_k-\phi_l)}\rangle\rangle=d\delta_{kl}
,\label{e-disorder} 
\ee
where the parameter $d\ll 1$ characterizes the disorder strength.
\end{subequations}

Our main result  is the mapping of the original Hamiltonian \rref{problem} 
to the well studied problem \cite{GS}
of  repulsive electrons in the disordered potential. The latter model is described by the Matsubara action
\begin{subequations}
\label{result}
\be
 \mathcal{S}
=  \int dx d\tau
\frac{\left(\partial_\tau \alpha\right)^2/u+
\left(\partial_x \alpha\right)^2 u}{2 \pi K}
+ \Re\,\varepsilon(x)e^{2i\alpha(x,\tau)}.
\label{result1}
\ee
Here $\alpha(x,\tau)$ is the  real bosonic field related to the electron density by
\be
\rho= (\bpsi^\dagger\bpsi) \to -\frac{\partial_x\alpha}{2\pi}.
\label{result2}
\ee

The complex field $\varepsilon(x)$ describes the static disorder 
\be
\langle\langle\varepsilon(x)\varepsilon^*(x')\rangle\rangle=
\frac{u^2}{\xi^3} {\cal D}\delta(x-x').
\label{result3}
\ee
Action \rref{result1} is analogous to that of the pinning of the charge density waves by the point like impurities
and the randomness of the phases of $\varepsilon(x)$ describes the randomness of the positions of such impurities with
respect to the charge density wave period.

Luttinger parameter $K$, sound velocity $u$, minimal length of the
validity of the effective description \rref{result1}, $\xi$, and the
dimensionless disorder strength ${\cal D}$ are related to the parameters of the original model \rref{problem}
\be
\begin{split}
K= \frac{4 u}{v};\ u=\frac{J}{2\pi}\left(\ln \frac{E_B}{\Delta}\right)^{1/2};
\ \xi=\frac{2av}{J};\
{\cal D}=\frac{dv}{J}\ln\frac{E_B}{\Delta},
\end{split}
\label{result4}
\ee
where $E_B$ is the width of the band of the chiral fermions,
and $\Delta= J/2a$ has the meaning of an averaged single particle gap.
The description is valid provided that the original coupling of the electrons with spins is weak enough $J\ll v$, {\em i.e.} $K\ll 1$ and $\xi \gg a$ (also the relative values of the couplings  $j_z^{k}$ are assumed to be moderate).
\end{subequations}

Mapping \rref{result} enables us to make several important conclusions about the low temperature transport.
Indeed, without  the anisotropy disorder, $d=0$, the current-current
correlator found from \reqs{result1}--\rref{result2} retains the ballistic pole structure
\be
\langle jj\rangle_{\Omega,q}=\frac{e^2\Omega^2\langle\rho\rho\rangle_{\Omega,q}}{q^2}
=
\frac{e^2 v}{\pi}\frac{ (K/4)^2\Omega^2}{\Omega^2+u^2q^2}
\label{jj}
\ee
indicating the absence of the Anderson localization.
Correlation function \rref{jj} gives the Kubo conductivity
$\sigma(\omega)=  
\langle jj\rangle_{\Omega\to -i\omega+0,q=0}/(-i\omega) \propto 1/\omega$.
The conductance of the infinite system \cite{FisherLee} $G=\int \frac{dq}{2\pi\hbar|\Omega|}
\langle jj\rangle_{\Omega,q}=e^2(K/4)/(2\pi\hbar)$.
Moreover, in analogy with Refs.~\cite{ballistic}, one may expect that the conductance of the finite system connected to ideal leads retains its universal value $e^2/(2\pi\hbar)$.

Arbitrary weak anisotropy disorder $d>0$ changes the picture drastically.
Indeed for the repulsively interacting electrons, the disorder is always
relevant perturbation: its evolution with the microscopic scale $\xi$
is given by \cite{GS}
\be
\frac{d{\cal D}}{d\ln \xi }=\left(3-2K\right){\cal D}.\label{RG1}
\ee
Combining \req{result4} with \req{RG1} we find the estimate for the localization
length of the chiral fermions, ${\cal D}(\xi=L_{loc}) \simeq 1$,
\be
L_{loc}=a \left(\frac{v}{J}\right)^{\frac{2-2K}{3-2K}}
\left(\frac{1}{d \ln(E_B/\Delta)}\right)^{\frac{1}{(3-2K)}}.
\label{localization-length}
\ee

{\em Semiclasical approximation for the spin dynamics}
 is valid because of the important properties of the effective spin interaction
in the chiral wire which we illustrate now.
In the second order in $J$, one obtains
\be
\hH_s=- \frac{J^2}{4\pi v}\sum_{k \neq l}
\frac{\hS_\alpha^k w_{\alpha\beta}^{kl}\hS_\beta^l}{|x_k-x_l|}.
\label{spin-coupling1}
\ee
The couplings $w_{\alpha\beta}^{kl}$ are related to the parameters of \req{couplings}
as
\be
\begin{split}
&  w_{\alpha\beta}^{kl}=
\left[
\hO^k\left(\hat{P}+\epsilon^k\hat{\Phi}^k+\epsilon^l\hat{\Phi}^l
\right)\hO^l \right]_{\alpha\beta};
\\
&\hat{p}=
\begin{pmatrix}1&0&0\\
0&1&0\\0&0&0\end{pmatrix}
;\ \ \hat{\Phi}^k=
\begin{pmatrix}\cos \phi^k&\sin\phi^k&0\\
\sin\phi^k&-\cos\phi^k&0\\0&0&0\end{pmatrix};
\end{split}
\label{w}
\ee
where only terms linear in disorder $\epsilon$ are kept.
We notice that unlike the case of non-chiral systems:
(1) coupling constants  $j_{x,y,z}$ do not enter into spin interaction
at all; (2) most of the disorder can be got rid of by the rotation
\be
\hS_\beta^k O^k_{\beta\alpha}\to \hS_\alpha,
\label{rotation1}
\ee
which leaves the commutation relations for the spin operators intact.
It means that for $\epsilon=0$ the system of rotated spins is classically
ferromagnetically ordered in $X-Y$ plane ($w_{xx}=w_{yy}=1,\ w_{zz}=0$),
and no frustration is possible.
For the same reason the induced spin-spin interaction always favor the collectively
ordered state rather than single impurities surrounded by the Kondo cloud.
 Therefore, the low-energy physics
of the system is determined by the collective motion of the rotated spins
\rref{rotation1} which effective description we develop.

We parameterize the spins by the smooth fields $\alpha(x,\tau)$ and $-1<n_z(x,\tau)<1$
\be
S^k_\alpha(\tau)=\frac{1}{2}\left[\hO^k
\begin{pmatrix}
\sqrt{1-n_z^2(x_k,\tau)}\cos\alpha(x_k,\tau)\\
 \sqrt{1-n_z^2(x_k,\tau)}\sin\alpha(x_k,\tau)\\
n_z(x_k,\tau)\end{pmatrix}\right]_\alpha
\label{smooth}
\ee
Then the system \rref{problem} has the Matsubara action representation
$\mathcal {S}=\mathcal {S}_{el}+\mathcal {S}_{s}$.
The spin part of the system is nothing but familiar \cite{Nagaosa}
Wess-Zumino action
\begin{subequations}
\label{action}
\be
\mathcal{S}_{s}=
{-i} \int d\tau dx \, \rho_s(x)\, n_{z}(x,\tau)\partial_{\tau}\alpha(x,\tau).
\label{WZ}
\ee
where $2 \rho_s(x)=\sum_{k}\delta(x-x_k)$ is the linear density of the localized spin
$2\langle\langle\rho_s\rangle\rangle=1/ a$.

The action for chiral electrons can be conveniently represented as
$\mathcal{S}=\mathcal{S}_0+\mathcal{S}_\perp+\mathcal{S}_\parallel$.
The term
\be
\begin{split}
&\mathcal{S}_0=
\int dxd\tau\bar{\bpsi}\left[\partial_\tau-iv\htau_z\partial_x
+{J\rho_s}\sqrt{1-n_z^2}\, \htau(\alpha)
\right]\bpsi;
\\
&\htau(\alpha)= \htau_x\cos\alpha+\htau_y\sin\alpha,
\end{split}
\raisetag{5mm}
\label{S0}
\ee
($\bar{\bpsi},\bpsi$ being two component Grassmann fields) describes the motion
of the electrons in the field created by  smooth and slowly evolving configurations
of the spins. 
Without spin dynamics the electron would be gapped and
no charge transport were possible. The spin dynamics in fact facilitate the transport
(similarly to the moving charge density wave).

The action $\mathcal{S}_\perp$ given by
\be
\mathcal{S}_\perp=
\int dxd\tau\bar{\bpsi}\left[{J\rho_s\epsilon}\sqrt{1-n_z^2}\,
\htau(-\alpha+\phi)
\right]\bpsi.
\label{Sperp}
\ee
contains the most relevant disorder. 
Here $\epsilon(x), \phi(x)$ are the random {\em static} fields such as
$\epsilon(x_k)=\epsilon^k,\ \phi(x_x)=\phi^k$.

The remaining term $\mathcal{S}_{\parallel}$ describes the forward scatterings by the spin.
It has the form
\be
\begin{split}
&\mathcal{S}_{\parallel}=
\int dxd \tau\bar{\bpsi}h_z(x,\tau)\htau_z\bpsi;\\
&h_z={J\rho_s}
\left[n_zj_z(x)+\sqrt{1-n_z^2}(j_x(x)\cos\alpha+j_y(x)\sin\alpha)\right],
\end{split}
\label{Sparallel}
\raisetag{14mm}
\ee
where once again
$j_\alpha(x)$ are the random {\em static} fields such as
$j_\alpha(x_k)=j^k_\alpha$.
 \end{subequations}

Let us turn to the analysis of the action \rref{action}. If there were no
fluctuation of $\rho_s,\ n_z$, and $\alpha$, the system \rref{S0} would acquire
the gap of the width $2\Delta=J/a$.
The fluctuations of $\rho_s,\ n_z$ are the static and dynamic
fluctuations of the gap, they  are massive and can be treated perturbatively
at low energies.
When fermions are gapped, the direct calculation of the fermionic determinants is more convenient than bosonization procedure  \cite{Book}.
Unlike the electrons the phase mode $\alpha$ remains soft and its
gradient must be taken into account. The most  convenient way to proceed is to
perform the gauge transformation
\begin{subequations}
\be
\bpsi=e^{-i\alpha\htau_z/2}\bpsi;
\ \bar{\bpsi}=\bar{\bpsi}e^{i\alpha\htau_z/2}.
\label{gautge-transformation}
\ee

As the result of such gauge transformation and recollection of terms
with diagonal and off-diagonal Pauli matrices 
the action $S_0$ becomes independent of $\alpha$
\be
\mathcal{S}_0=
\int dxd\tau\bar{\bpsi}\left[\partial_\tau-iv\htau_z\partial_x
+{J\rho_s}\sqrt{1-n_z^2}\, \htau_x
\right]\bpsi;
\label{S0gt}
\ee
term
 $\mathcal{S}_\perp$ remains $\alpha$ dependent
 \be
\mathcal{S}_\perp=
\int dxd\tau\bar{\bpsi}\left[{J\rho_s\epsilon}\sqrt{1-n_z^2}\,
\htau(-2\alpha+\phi)
\right]\bpsi,
\label{Sperpgt}
\ee
and \req{Sparallel} modifies as
\be
\begin{split}
\mathcal{S}_{\parallel}=-\frac{1}{2}
\int dxd \tau\bar{\bpsi}
\left[v\partial_x\alpha+i\htau_z(\partial_\tau\alpha+2ih_z)\right]\bpsi,
\end{split}
\label{Sparallelgt}
\ee

The spin part of the action \rref{WZ} acquires the additional (anomalous) part
$\simeq (\partial_x\alpha)^2$
\be
\mathcal{S}_{s}=
\int d\tau dx \,\left[-{i} \rho_s n_{z}\partial_{\tau}\alpha+
\frac{v(\partial_x\alpha)^2}{8\pi}
\right].
\label{WZgt}
\ee
\label{g-t}
The same anomaly is responsible for the appearance of the $\partial_x\alpha$ term in \req{result2}.
\end{subequations}

One can consider terms \rref{Sperpgt} -- \rref{WZgt} as small corrections to the main term \rref{S0gt}. Let us
first assume that the fields $\rho_s(x)$ and $n_z(x,\tau)$
are smooth on the time scale $1/\Delta$ and on the linear scale $\xi=v/\Delta$.
Under this assumption the fermions can be integrated out and we obtain
\begin{subequations}
\be
\mathcal{S}_0 \to - \frac{1}{2\pi v}
\int dx d\tau {J^2\rho_s^2}\left(1-n_z^2\right)
\ln \left(\frac{E_B}{J\rho_s\sqrt{1-n_z^2}}\right),
\label{nofermions1}
\ee
(as the system is gapped, non-anomalous contributions vanish from the density operator
\rref{result2}).
This action is maximized for $n_z=0$ and we will consider the small deviations from this extremum value \cite{footnote}.
The r.m.s. of static fluctuations of the spin density 
$\delta \rho_s=\rho_s-(1/2a)$ on the scale $\xi$ is suppressed by the the parameter $(a/\xi)^{1/2} \ll 1$.
It permits to neglect the fluctuation in the density under the logarithm and we obtain  in the leading logarithmic approximation
\be
\mathcal{S}_0+\mathcal{S}_s + \mathcal{S}_{\parallel} \rightarrow - \int d\tau dx
\frac{J^2\rho_s^2}{2\pi v}
\ln \left(\frac{E_B}{J\rho_s}\right)
+ \bar{\mathcal{S}} + \delta\mathcal{S}_Z;
\label{nofermions2}
\ee
The first term is nothing but sample-to-sample fluctuating ground state energy which can be neglected. The second term describes the soft mode which propagates
ballistically (at $h_z=0$)
\be
\bar{\mathcal{S}}=
\int d\tau dx \,\left[\frac{2\pi u^2\tilde{\rho}^2}{v}
-{i} \tilde{\rho}\partial_{\tau}\alpha+
\frac{v(\partial_{x}\alpha)^2}{8\pi }
\right]
,
\label{nofermions3}
\ee
with velocity $u \ll v$ defined in \req{result4}. We introduced the short hand notation $\tilde{\rho}(x,\tau)\equiv\rho_s(x)n_z(x,\tau)$. 
Notice, that the disorder disappeared from \req{nofermions3} completely.
Action \rref{nofermions3} is quadratic in $\tilde{\rho}$ and the partition function is determined solely by the saddle point
\be
\tilde{\rho}= \frac{i v\partial_\tau\alpha}{4\pi u^2},
\label{nofermions4}
\ee
substitution of this solution into \req{nofermions3} and the condition $u\ll v$ yields the first term in \req{result1}.

Remaining terms are locally small and have to be calculated for fields constrained by  \req{nofermions4}.
The contribution $\delta \mathcal{S}_Z$
is generated by the term \req{Sparallelgt} in the second order and it reduces to
\be
\begin{split}
\delta\mathcal{S}_Z&=\frac{1}{8\pi v}
\int dx d\tau\left(\partial_\tau\alpha+2ih_z\right)^2
\\
= &\frac{1}{8\pi v}
\int dx d\tau
\left(\partial_\tau\alpha +iJ j_z\tilde{\rho}\right)^2
\\
-  &\frac{J^2}{2\pi v}\int dx d\tau \rho^2_s(j_x\cos\alpha+j_y\sin\alpha)^2
,
\end{split}
\label{nofermions5}
\ee
where we omitted the total time derivatives of
the periodic functions of $\alpha$.
The  term in the
second line contains the small (as $J/v$) correction  to the velocity $u$.
As $j_z \lesssim 1$ it has to be neglected (even though this correction is a random function) as the stiffness disorder is irrelevant for low energy waves.
The last term has the same structure (terms $\mathrm{e}^{\pm 2i\alpha}$) 
as the last term in \req{result1}. However for $\langle\langle\epsilon^2\rangle\rangle \simeq \langle\langle j^2_{x,y} \rangle\rangle$, the absolute value of this term is much smaller that the contribution we will consider next.

The main contribution leading to the Anderson localization comes from \req{Sperpgt}.
Putting $n_z=0$ and averaging over fermions using \req{S0gt},
we obtain
\be
\mathcal{S}_\perp \rightarrow
\int dxd\tau
\frac{2\pi u^2}{a v}\left(\rho_s\epsilon\right)
\cos(2\alpha-\phi).
\ee
Introducing complex field
\be
\varepsilon(x)=\frac{2\pi u^2 \rho_s(x)\epsilon(x)}{av}e^{-i\phi(x)},
\label{nofermions6}
\ee
we arrive at \reqs{result1} and \rref{result3}.


 \label{gtnofermions}
\end{subequations}

{\em To conclude}
--
We analyzed the chiral electrons on an edge of a two-dimensional topological insulator allowing for all possible couplings with localized spins.
We showed that the system can be always reduced to the effective bosonic model \rref{result} which exhibits the Anderson localization.
In our explicit analysis we did not account for any interaction between the edge electrons.
However, we believe that our conclusions remain valid for any interaction of a finite strength. 
Indeed, as the effective bosonic model \rref{result} corresponds to the model with  Luttinger parameter $K \ll 1$ the finite electron-electron interaction can 
lead to only perturbative corrections and
is unable to destroy the Anderson localization. Only if the interaction is strongly attractive (which does not seem to be realistic)
can the localization be lifted and the superconductor-insulator transition \cite{GS} may occur.

We are grateful to  V. Cheianov, L.Glazman, A. Tsvelik, and A. Yacoby for helpful discussions and
to INT of University of Washington, where this work was completed.
This work was supported in part by NSF-CCF Award 1017244 (B.A.), Simons foundation (I.A.), and RFBR grant 12-02-00100 (V.Y.).


\begin{references}
\bibitem{Other}
Effects of crystal anisotropy specific for large impurity spins and
effects of hyperfine interactions with nuclear spins were
considered in Refs.~\onlinecite{GlazmanCh} and \onlinecite{NSpin} respectively.
\bibitem{Parametrization} For $\hat{J} = J[\mathrm{diag}(1, 1, \tilde{J}_z) + 
\delta\hat{J}]$ with $||\delta\hat{J}|| \ll 1$ the elements of the matrix $\hO = 1 - \hat{A}$ ($\hat{A}$ is a small antisymmetric matrix) and other parameters in Eq.(\ref{couplings}) are determined as $A_{xy} = (\delta J_{yx} - \delta J_{xy})/2$, 
$A_{xz} = \delta J_{zx}$, $A_{yz} = \delta J_{zy}$; $\epsilon\cos{\phi} = (\delta J_{xx}-\delta J_{yy})/2$, $\epsilon\sin{\phi} = (\delta J_{yx}+\delta J_{xy})/2$, etc.
\bibitem{footnote} This approximation neglects: i) quantum phase slips
in $\alpha(x,\tau)$ requiring $n_z=\pm 1$ at points of the phase slips;
(ii) exponentially rare fluctuations in the density of localized spins.
Phase slips are strongly irrelevant at $K\ll 1 $, and the rare fluctuations
are not important for large localization length $L_{loc} \gg a$, so both
 assumptions are justifiable.
\end{references}
\end{document}